\documentclass[reprint,superscriptaddress]{revtex4-1}
\usepackage{amsmath,graphicx}

\begin{document}

\title{Micromechanical Microphone using Sideband Modulation of Nonlinear Resonators}
\date{30 August 2017}

\author{Joseph A. Boales}
\affiliation{Department of Physics, Boston University, 590 Commonwealth Avenue, Boston, MA 02215, US}
\author{Farrukh Mateen}
\affiliation{Department of Mechanical and Aerospace Engineering, Boston University, 110 Cummington Street, Boston, MA 02215, USA}
\author{Pritiraj Mohanty}
\affiliation{Department of Physics, Boston University, 590 Commonwealth Avenue, Boston, MA 02215, US}

\begin{abstract}
We report successful detection of an audio signal via sideband modulation of a nonlinear piezoelectric micromechanical resonator. The 270-by-96-$\mu$m resonator was shown to be reliable in audio detection for sound intensity levels as low as ambient room noise and to have an unamplified sensitivity of 23.9 $\mu$V/Pa. Such an approach may be adapted in acoustic sensors and microphones for consumer electronics or medical equipment such as hearing aids.
\end{abstract}
\maketitle

The first microphone was invented and patented by Emile Berliner in the late nineteenth century \cite{berliner1880microphone}. Since then, microphone diaphragm sizes have shrunken to astounding sizes and continue to shrink \cite{scheeper1994review,lee2008piezoelectric,li2017zno}. One constant motivation for the size reduction is to fit more functionality into modern smartphones and other smart devices while maintaining a small form factor; however, smaller microphone diaphragms are not without their drawbacks. Namely, the size of the diaphragm largely controls its signal-to-noise ratio \cite{bae2004design,tanaka2007industrial,mohd2009noise}, which can dramatically impact the usability of devices such as hearing aids \cite{levitt2001noise}.

Countless microphone types for applications ranging from voice recording to medical ultrasound have been realized, each requiring its own special design \cite{scheeper1994review,ballantine1996acoustic}. As recently as 2009, a new type of laser-based microphone, which measures the vibrations of particulates suspended in air, was patented \cite{schwartz2009particulate}. However, this technology is cumbersome and expensive. For practical applications, piezoelectric micromechanical (MEMS) microphones have become an area of intense interest \cite{kressmann1996new,schellin1992silicon,johnson2014startup,littrell2010high}. Microfabricated MEMS microphones can be produced with astoundingly small form factors and be built directly into semiconductor chips \cite{scheeper1994review,williams2012aln}.

Here, by taking advantage of the frequency-mixing properties of nonlinear MEMS resonators constructed from a combination of silicon structure and aluminum nitride active layer, we have successfully and reliably detected sound waves using a device with a top surface area of only $2.6\times10^{-8}\ \mathrm{m^2}$. Sound intensity levels as low as 54 dBA were detectable using this device. Our setup can be used to produce a microphone with sensitivity comparable to current state-of-the-art devices \cite{williams2012aln}.

As we have demonstrated in previous work \cite{boales2017optical}, it is possible to transmit information by applying a small, off-resonance, time-varying force to a nonlinear mechanical resonator that is being strongly driven at one of its resonance frequencies. In our previous work, we demonstrated this using optical radiation pressure in vacuum. This had the advantage of increasing the quality factor of the resonator by removing losses due to air \cite{mohd2009noise,imboden2014dissipation,vig1999noise}.

In this paper, we present the results of a similar experiment that uses acoustic pressure waves as the small signal rather than modulated optical radiation pressure. In contrast to the previous experiment, the acoustic pressure wave inherently requires a medium for propagation. Despite the much higher damping and lower quality factor that is present when the resonator is exposed to air, we were able to consistently detect sound waves with high sensitivity.

We have previously shown that a nonlinear response for a single vibration mode of the resonators in this experiment can be modeled using the equation \cite{imboden2014dissipation}
\begin{align}
m \ddot{x} + \gamma \dot{x} + k x &+ k_3 x^3 \nonumber \\
&= A_r \cos(2\pi f_r t) + A_m \cos(2 \pi f_m t)
\end{align}
where $m$ is the effective modal mass, $\gamma$ is the linear damping factor, $k$ is the effective modal spring constant, $k_3$ is the cubic nonlinear spring constant, $A_r$ is the resonance driving amplitude, $A_m$ is the driving amplitude produced by the sound waves, $f_r$ is resonance frequency, $t$ is time, and $f_m$ is the frequency of the sound wave. In this experiment, $A_m$ is proportional to $PA$, where $P$ is the amplitude of the pressure wave and $A$ is the effective modal area of the resonance mode.

The pressure wave used in this basic analysis is also known as the Langevin (rather than Rayleigh) acoustic radiation pressure, which is the average difference between the force per area applied to the front surface of the device and the ambient pressure at the back surface \cite{beyer1978radiation}. More recently, it has been shown that Rayleigh acoustic radiation pressure is the acoustic radiation pressure that acts on a moving surface, while the Langevin radiation pressure acts on a stationary surface \cite{hasegawa2001general}. In this experiment, the vibrational frequency of the resonator is orders of magnitude larger than the acoustic frequency, so the resonator is at rest on average over the period of the acoustic wave. The amplitude of mechanical oscillation is also negligible. For both of these reasons, the resonator can be treated as being stationary.

A steady-state solution to equation (1) near the first harmonic of the resonance can be written as \cite{boales2017optical}
\begin{align}
x(t) \approx c_r \cos(&2 \pi f_r t) +  c_m \cos(2 \pi f_m t) \nonumber \\
&+ \sum\limits_{n} c_3(n) \cos\left(2 \pi (f_r \pm n f_m) t \right)
\end{align}
where $c_r$, $c_m$, and $c_3$ are response amplitudes and $n$ is a positive integer. In the absence of nonlinearity, $c_3(n)$ is zero for all $n$. Frequency and amplitude information contained within the modulation signal can be decoded by demodulating the sidebands at frequencies $f_r \pm nf_m$. For convenience, we use the first-order upper sideband, which is located at the frequency $f_r+f_m$, in this experiment.

\begin{figure*}
\includegraphics[width=0.8\textwidth]{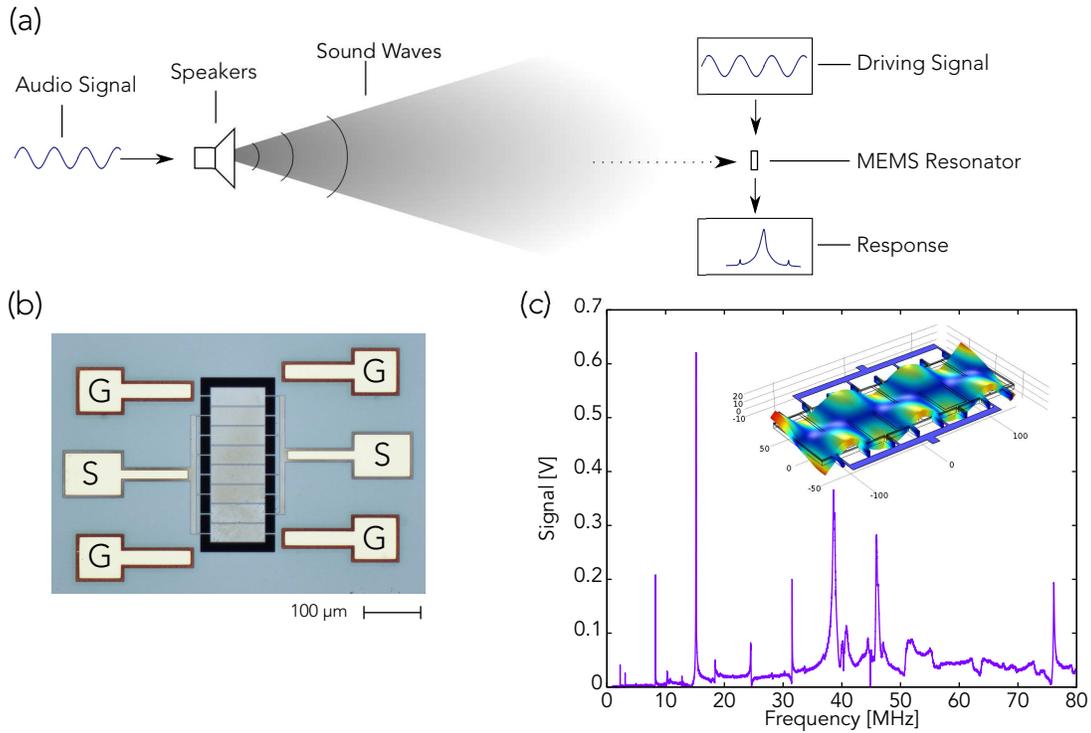}
\caption{(a) Experimental setup. A piezoelectric MEMS resonator is electrically driven using a signal generator and its response is measured using a spectrum analyzer. At the same time, an audio signal is provided to speakers, and the sound wave they produce is incident on the resonator. (b) Micrograph of the resonator used for this experiment. Electrodes marked ``S" are used for driving or measuring the response of the resonator. Electrodes marked ``G" are used to access the grounding plane. (c) Frequency spectrum of resonator the resonator when driven at 19 dBm in the range 1 to 80 MHz.The inset is the modeshape of the 15.168 MHz mode as generated by COMSOL.}
\label{fig:Figure1}
\end{figure*}

As illustrated in Figure \ref{fig:Figure1}(a), we electrically drive a piezoelectric MEMS resonator at resonance using a signal generator at 19 dBm. An audio speaker is placed at a fixed distance from the resonator and provided a signal at a single frequency, typically 200 Hz. The resonator's response is amplified, then measured using a spectrum analyzer. As shown in the micrograph in Figure \ref{fig:Figure1}(b), the resonator is a 270-by-96-$\mu$m rectangular plate which is suspended by sixteen 15-by-3-$\mu$m legs. From bottom layer to top layer, it is constructed from a 5-$\mu$m silicon and 1-$\mu$m silicon dioxide structure, a 300-nm molybdenum ground plane, 1-$\mu$m aluminum nitride (AlN) piezoelectric layer, and 300-nm interdigitated molybdenum electrodes. Signals can be electrically measured or applied at the electrodes labeled ``S", and the electrodes marked ``G" provide access to the ground plane. The resonator is directly electrically driven via the inverse piezoelectric effect, where a potential applied across the AlN causes a strain in the layer. The response is measured via the direct piezoelectric effect, where a strain in the AlN layer produces a potential difference between the molybdenum layers.

The resonator contains a number of resonant modes in the frequency range of 1 to 80 MHz, as shown in Figure \ref{fig:Figure1}(c), the most prominent one being the 15.168 MHz mode. The mode shape, generated using COMSOL Multiphysics, is shown in the inset of Figure \ref{fig:Figure1}(c).  The mode shape and frequency can be similarly approximated by solving the Euler-Bernoulli equation. This resonance mode is used for the remainder of the experiment. For this mode, $m$ is approximately 57.7 ng, $k$ is 523.8 kN/m, and $\gamma$ is $6.10\times10^{-6}$ Ns/m.

\begin{figure*}
\includegraphics[width = 0.8\textwidth]{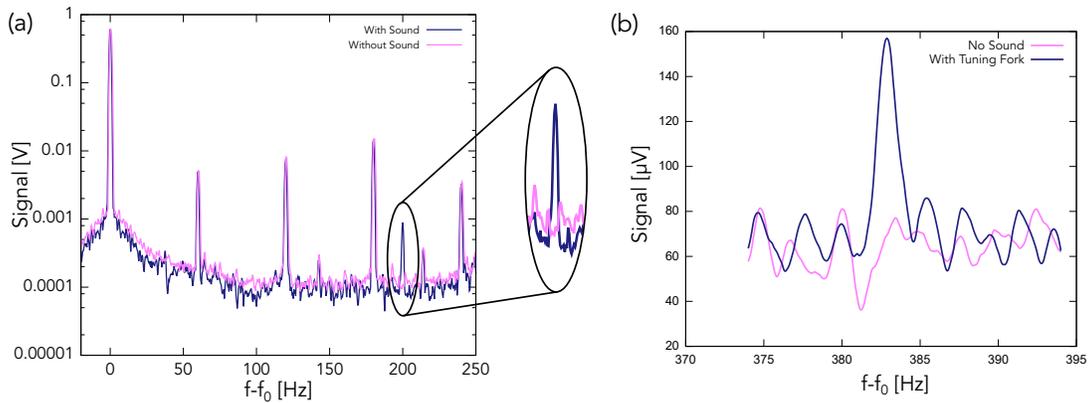}
\caption{(a) Response of resonator when driven with power 19 dBm at 15.168 MHz. The pink line was recorded with only ambient noise present, and the blue line was recorded while the speaker was producing a 200 Hz audio tone. (b) Response of resonator in range 374 to 394 Hz above resonance when driven with power 19 dBm at 15.168 MHz. The pink line was recorded with only ambient noise present, and the blue line was recorded while a musical tuning fork was making sound. This data was collected to rule out electronic noise as the source of the produced sideband.}
\label{fig:Figure2}
\end{figure*}

Next, we drove the resonator at 15.168 MHz and measured its response, shown by the pink line in Figure \ref{fig:Figure2}(a). The resonance peak is the furthest to the left, and the other peaks are primarily due to 60 Hz noise sources and internal instrument noise. The x-axis shows the frequency relative to the driving frequency. We then turned on the speaker and measured the response again, as shown by the blue line in Figure \ref{fig:Figure2}(a). With the speaker turned on, a prominent peak appeared at 15.1682 MHz, 200 Hz above the driving frequency. The magnified oval on the plot shows that, with the speaker turned off, there is no peak present at that frequency.

As evident from the large number of peaks in Figure \ref{fig:Figure2}(a), electronic noise is also a potential source that can produce sidebands. To rule out electronic noise as the source of the sideband observed during the 200 Hz speaker experiment, we used a 384 Hz mechanical tuning fork (such as those used for tuning musical instruments) to produce a sideband 384 Hz above resonance, as shown in Figure \ref{fig:Figure2}(b). When the tuning fork is making sound, a sideband peak that is approximately 100 $\mu$V larger than the background is present. When it is silent, the sideband is not present. The tuning fork has the advantage that it is a purely mechanical source of acoustic waves, so electronic interference is not possible; hence, the sideband must be a result of acoustic pressure waves.

\begin{figure*}
\includegraphics[width = 0.8\textwidth]{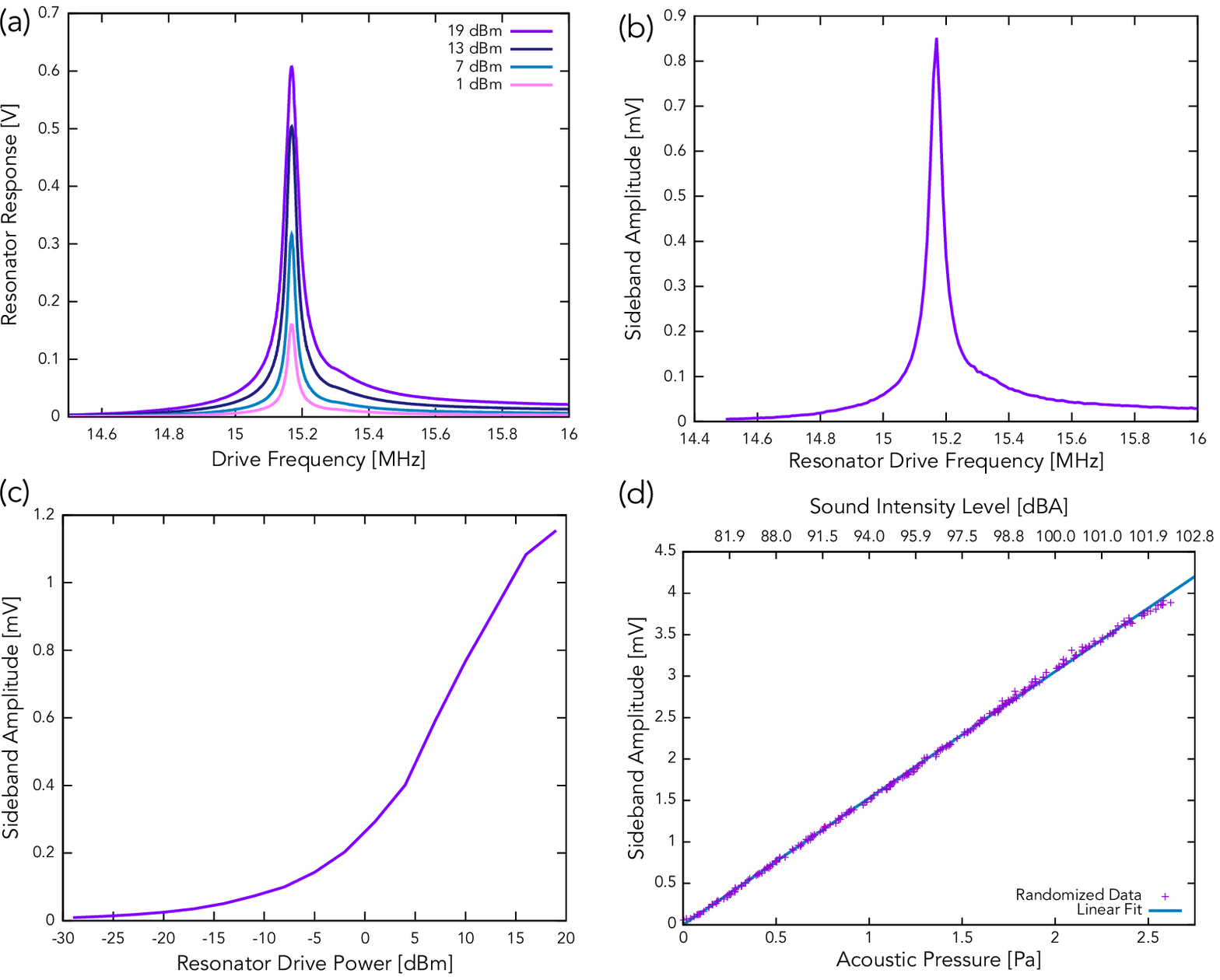}
\caption{(a) Response of resonator between 14.5 and 16.0 MHz when directly driven at various powers. (b) Amplitude of the first order upper sideband when the resonator is driven with power 19 dBm between frequencies 14.5 and 16.0 MHz. (c) Sideband amplitude as function of resonator driving power while resonator is driven at 15.168 MHz. (d) Sideband amplitude as function of sound intensity. Resonator is driven at constant power 15.168 MHz with power 19 dBm and acoustic wave is applied at randomly generated amplitudes.}
\label{fig:Figure3}
\end{figure*}

After verifying that the sideband was indeed a result of the sound produced by the speaker, we further characterized the resonator and the sideband for various operational parameters. Figure \ref{fig:Figure3}(a) shows the shape of the resonance peak as a function of frequency and for several different driving powers. Next, we measured the size of the first-order upper sideband as a function of driving frequency, Figure \ref{fig:Figure3}(b). For this plot, a 200-Hz sound wave was continuously incident on the resonator, and the sideband amplitude was measured as the frequency of the 19-dBm signal provided to one of the “S” terminals was varied. By comparing Figures \ref{fig:Figure3}(a) and \ref{fig:Figure3}(b), it is clear that the sideband amplitude is directly related to the resonant response amplitude of the resonator, as expected.

Furthermore, we drove the resonator at 15.168 MHz and measured the sideband amplitude while varying the resonator drive power from 19 dBm to -29 dBm in 3 dBm increments, Figure \ref{fig:Figure3}(c). Once again, this demonstrated that the sideband amplitude is directly proportional to the resonance response amplitude. Finally, we measured the dependence of the sideband amplitude on the sound level intensity, Figure \ref{fig:Figure3}(d). For this measurement, an acoustic wave with RMS pressure between 0 and 2.7 Pa was produced by the speakers. The sound level intensity was calibrated using a standard sound level meter (Protmex MS6708).  As shown in Figure \ref{fig:Figure3}(d), the sideband amplitude is directly proportional to the amplitude of the pressure wave applied, and hence the size of the force applied by the acoustic wave. This linearity is consistent with our predictions and with results of previous published works \cite{boales2017optical}. Using this data, we find that, including our preamplifier, these devices can be used as microphones with a sensitivity of 1.53 mV/Pa. Without signal amplification, the sensitivity is 23.9 $\mu$V/Pa.  For the data in Figure \ref{fig:Figure3}(a), the spectrum analyzer was set to a bandwidth of 9.1 kHz and set to hold its maximum value. For the remaining subfigures, the spectrum analyzer was set to a bandwidth of 2 Hz and averaged 10 times for each measurement.

Recently, a design for an AlN-based MEMS microphone was demonstrated to have a sensitivity comparable to the results presented in this paper \cite{williams2012aln}. However, unlike the previously published results, our microphone takes advantage of the nonlinear nature of MEMS resonators. While other state-of-the-art microphones have been shown to have sensitivities in the range of 200 $\mu$V/Pa \cite{kressmann1996new,schellin1992silicon,littrell2010high}, they have not taken advantage of the mode-mixing properties of nonlinear MEMS devices. The sensitivity produced by our method can be further enhanced by using improved or specially-designed resonator shapes.

It is important to note that the resonator design and equipment used for this demonstration are intended only as a proof-of-concept; the resonators have not been optimized for this application, nor has the measurement equipment used been miniaturized. Future work in this project includes optimizing the design for both increased sensitivity and an increased signal-to-noise ratio. Further, we acknowledge that operation in the nonlinear regime has the marked disadvantage of increased power consumption. For instance, throughout this experiment, we used a power of 79 mW to drive the resonator. Resonators which may operate in their linear regimes require only tens or hundreds of microWatts. An easily available MEMS microphone from Analog Devices (Model ADMP401) has a sensitivity of -42 dBV and requires only 0.83 mW of power. While it is certainly possible to operate near the linear regime, high powers provide a marked improvement in sensitivity, as demonstrated in Figure \ref{fig:Figure3}(c). However, optimization of our nonlinear microphone design may enable their use for measurement of ultrasensitive signals where conventional linear microphones are impractical. For instance, the Analog Devices package is 4.72 mm $\times$ 3.76 mm, requiring more than 680 times more surface area than the resonators from this experiment \cite{analogdevices:ADMP401}. In order to further compete with existing devices, improved piezoelectrics such as Sc-AlN can be used to improve signal transduction compared to AlN \cite{umeda2013piezoelectric} and structural materials such as diamond may be used to improve the quality factor of the resonators \cite{bautze2014superconducting}.

In conclusion, we have shown that a piezoelectric MEMS resonator can easily be driven strongly enough in air to display nonlinear behavior, which can be used to detect audio signals as small as 54 dBA with an effective diaphragm size as small as $2.6\times10^{-8}\ \mathrm{m^2}$. The sensitivity of this device was further shown to be 23.9 $\mu$V/Pa, or -92.4 dBV. For practical applications such as audio microphones and hearing aids, similar resonators may be arrayed and used in conjunction with modern demodulation methods to rival commercially-available state-of-the-art  microphones. 

\bibliography{references}

\end{document}